# Binaural Sound Event Localization and Detection based on HRTF Cues for Humanoid Robots

Gyeong-Tae Lee, Hyeonuk Nam, and Yong-Hwa Park

*Abstract*— This paper introduces Binaural Sound Event Localization and Detection (BiSELD), a task that aims to jointly detect and localize multiple sound events using binaural audio, inspired by the spatial hearing mechanism of humans. To support this task, we present a synthetic benchmark dataset, called the Binaural Set, which simulates realistic auditory scenes using measured head-related transfer functions (HRTFs) and diverse sound events. To effectively address the BiSELD task, we propose a new input feature representation called the Binaural Time-Frequency Feature (BTFF), which encodes interaural time difference (ITD), interaural level difference (ILD), and high-frequency spectral cues (SC) from binaural signals. BTFF is composed of eight channels, including left and right mel-spectrograms, velocity-maps, SC-maps, and ITD-/ILD-maps, designed to cover different spatial cues across frequency bands and spatial axes. A CRNN-based model, BiSELDnet, is then developed to learn both spectro-temporal patterns and HRTF-based localization cues from BTFF. Experiments on the Binaural Set show that each BTFF sub-feature enhances task performance: V-map improves detection, ITD-/ILD-maps enable accurate horizontal localization, and SC-map captures vertical spatial cues. The final system achieves a SELD error of 0.110 with 87.1% F-score and 4.4° localization error, demonstrating the effectiveness of the proposed framework in mimicking human-like auditory perception.

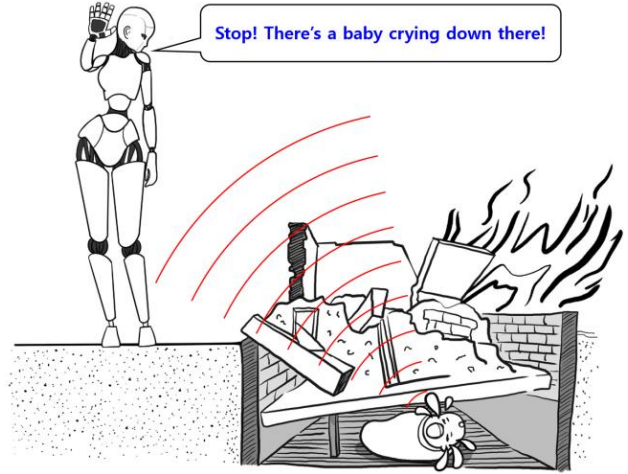

**Fig. 1.** Out of sight scenario of a disaster site.

## I. INTRODUCTION

Humanoid robots, designed to operate in human-centric environments, require not only anthropomorphic movement but also human-like perceptual abilities [1], [2]. While early research focused on bipedal locomotion (e.g., ASIMO, HUBO) [3]-[10], recent trends have shifted toward sensor-based intelligence, including vision and auditory perception, to enhance human-robot interaction. To effectively interact with humans, humanoid robots must perceive their surroundings in a multimodal manner. While early systems relied primarily on visual input via actuated cameras [11], subsequent developments incorporated other senses such as audition [12], olfaction [13], and gustation [14], recognizing their complementary roles in human-robot interaction [15]. Among them, audition has become a core modality, enabling speech recognition, environmental awareness, and social engagement [10]. The integration of vision and audition began as early as 1984 with WABOT-2 [12], and recent robots have demonstrated basic conversational abilities [10]. However, to respond contextually, robots must go beyond speech to recognize a wide range of real-world sound events.

Auditory perception enables humanoid robots to identify and localize sound sources beyond their visual field, providing essential situational awareness for operating in dynamic environments. For instance, recognizing alarms or sirens allows robots to avoid hazards or alert nearby individuals. More critically, in scenarios such as human rescue, auditory cues can compensate for occluded visual information. As illustrated in Fig. 1, detecting the cries of a baby trapped under debris requires the robot to interpret binaural signals from its microphones, highlighting the vital role of sound event localization and detection (SELD) capabilities in real-world deployment.

The main objective of this research is to develop a two-channel SELD framework tailored for humanoid robots [16]. Conventional horizontal two-channel input systems struggle with front-back confusion and elevation estimation due to the lack of spatial cues. Inspired by how humans overcome these challenges using head-related transfer functions (HRTFs), we assume a humanoid robot equipped with human-like ears—such as an android robot—would receive binaural signals embedded with its own HRTF characteristics. These cues must be extracted from the input signals to estimate the direction of arrival (DOA) of each sound source. Ideally, such a robot would internalize its own HRTF through learning, enabling it to detect and localize surrounding sound events simultaneously—just like humans. Building on this concept, we propose BiSELD

---

This paragraph of the first footnote will contain the date on which you submitted your paper for review, which is populated by IEEE. "This work was supported in part by the U.S. Department of Commerce under Grant 123456." *(Corresponding author: Yong-Hwa Park).*

Gyeong-Tae Lee, Hyeonuk Nam and Yong-Hwa Park were with Korea Advanced Institute of Science and Technology, Daejeon, South Korea

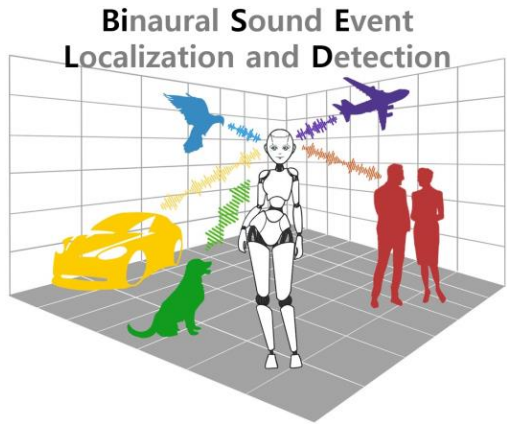

**Fig. 2.** Abstract concept of binaural sound event localization and detection (BiSELD) for a humanoid robot.

(Binaural Sound Event Localization and Detection), a task to jointly predict the class and direction of sound events from binaural audio. The core of our approach lies in a binaural time-frequency feature (BTFF), which encodes HRTF-based spatial patterns of sound events and serves as input to a deep learning model as illustrated in Fig. 2. The BiSELD task requires to detect target sound event classes while estimating the 3D direction of each event using only two-channel binaural input. To achieve this, we propose a CRNN-based baseline model, BiSELDnet.

## II. RELATED WORKS

### A. Computational Auditory Scene Analysis (CASA)

The concept of auditory scene analysis (ASA), first introduced by Bregman [17], describes how the auditory system decomposes complex acoustic mixtures into perceptually meaningful sound sources. Despite the limited information embedded in sound signals, humans reliably parse them with minimal confusion. This perceptual capability has been modeled computationally as computational auditory scene analysis (CASA) [18], which aims to emulate human auditory grouping and stream formation. CASA typically involves two stages: (1) grouping sound components based on Gestalt principles, and (2) selecting dominant streams through competitive mechanisms, resulting in coherent auditory streams that support source tracking and prediction [19].

While Bregman's ASA framework provided foundational insights through behavioral experiments, its neural and computational underpinnings remain only partially understood. To address this, various CASA models have been proposed and are typically classified into Bayesian approaches [20], [21], neural models [22]–[25], and temporal coherence-based methods [26], [27]. However, most focus on specific components of ASA, lacking integration into a unified model. Some models are inspired by hypothesized neural mechanisms or supported by neuroimaging evidence [19], yet the biological basis for auditory stream segregation remains incomplete. Consequently, despite their conceptual relevance, existing CASA models are not yet directly applicable to sound event detection and localization in humanoid robots.

### B. Sound Event Localization and Detection (SELD)

Sound event localization and detection (SELD) aims to jointly detect sound event activities and localize their spatial directions when active [28]. It comprises two sub-tasks: sound event detection (SED) and sound source localization (SSL). SELD plays a critical role in applications such as robotic navigation [29]–[31], audio surveillance [32], [33], biodiversity monitoring [34], and context-aware systems [35]. Since its inclusion in the DCASE Challenge Task 3 in 2019 [36], the field has progressed rapidly. Adavanne et al. [27] introduced SELDnet, a CRNN-based model combining convolutional layers, bidirectional GRUs, and fully connected layers to jointly estimate the class and direction of sound events. It served as the initial DCASE baseline and has since inspired a wide range of SELD architectures.

Since the introduction of SELDnet, numerous models have enhanced its architecture and input representations. Several works replaced the BiGRU with LSTM or temporal convolutional networks (TCNs) to improve temporal modeling [37], [38], [39]. Others introduced non-square or 1D convolutional filters to better capture feature dependencies [40], [41]. Input features have also evolved significantly. Many models combined log-mel spectrograms with spatial cues such as GCC-PHAT or intensity vectors to better inform the direction-of-arrival (DOA) estimation [41]–[43], [44]. Some models adopted dual-branch CRNNs for SED and SSL, utilizing techniques such as soft parameter sharing [45] or sequence matching [46]. A major milestone was the introduction of activity-coupled Cartesian DOA (ACCDOA) [47], which unified SED and DOA outputs into a single representation and loss function. It was further extended into multi-ACCDOA with auxiliary permutation invariant training (PIT) to handle overlapping events of the same class [48]. This representation has since been widely adopted in modern SELD systems. Additionally, SALSA [49] provided a time-frequency aligned spatial cue representation, supporting both FOA and MIC formats, though it introduced computational overhead. A simplified variant, SALSA-Lite, was later proposed for practical deployment in microphone array settings [50].

As discussed above, most SELD systems rely on multichannel input formats such as FOA and MIC, typically using four audio channels as per DCASE Challenge standards. While multichannel setups provide improved spatial resolution, they introduce significant trade-offs in terms of system complexity, data volume, and maintenance overhead. Specifically, four-channel input increases the burden on storage, transmission, and computation. Ensuring accurate spatial alignment requires regular calibration, and the physical form factor of microphone arrays can be unsuitable for compact or embedded applications. These limitations highlight the need for lightweight, two-channel SELD approaches, particularly in resource-constrained or form-factor-sensitive scenarios.

### C. Binaural SED, SSL, and SELD

Binaural audio input has recently gained attention in deep learning-based SED, SSL, and SELD due to its ability to replicate human auditory spatial cues. Using only two microphones, binaural signals encode essential localization information, including interaural time difference (ITD), interaural level difference (ILD), interaural phase difference (IPD), and spectral cues (SC). These features allow SELD systems to estimate 3D sound directions with minimal hardware complexity. In the context of humanoid robots, binaural approaches are particularly suitable for real-world deployment, offering a balance between spatial accuracy and system simplicity. A summary of recent deep learning-based binaural SED, SSL, and SELD methods is provided in Table I.

TABLE I
SUMMARY OF DEEP LEARNING-BASED BINAURAL SED, SSL, AND SELD METHODS.

| | Author | Architecture | Input feature | Output format | | |
|---|---|---|---|---|---|---|
| | | | | # Classes | Azimuth | Elevation |
| Binaural SED | Adavanne and Virtanen [51] | CRNN | Mel-spectrogram, Magnitude + Phase | 6 | − | − |
| | Krause and Mesaros [52] | CRNN | Mel-spectrogram, ILD, Phase, IPD, sin(IPD) + cos(IPD), GCC-PHAT | 62 | - | − |
| Binaural SSL | Youssef et al. [53] | ANN | ITD, ILD | − | −45° ~ +45° | × |
| | Roden et al. [54] | DNN | ITD, ILD, Magnitude + Phase, Real + Imaginary | − | −30° ~ +30° | −10° ~ +50° |
| | Nguyen et al. [55] | CNN | ILD + IPD | − | −45° ~ +45° | −30° ~ +30° |
| | Pang et al. [56] | CNN | ILD + IPD | − | −80° ~ +80° | −45° ~ +230.625° |
| | Yang et al. [57] | CNN | IPD, Log-magnitude | − | −80° ~ +80° | −45° ~ +230.625° |
| | Zermini et al. [58] | DNN | Mixing vector + ILD + IPD | − | −90° ~ +90° | − |
| | Ma et al. [59] | DNN | ILD, Cross-correlation | − | 0° ~ 360° | − |
| | Yang et al. [60] | CRNN | Log-magnitude, Phase | − | −80° ~ +80° | − |
| | García-Barrios et al. [61] | CRNN | Mean magnitude + sin(IPD) + cos(IPD) + ILD, Quaternions | − | −180° ~ +180° | −35° ~ +35° |
| | Dwivedi et al. [62] | CNN | Linear prediction residual coefficients | − | − | −30° ~ +210° |
| | van der Heijden and Mehrkanoon [63] | CNN | Bilateral auditory nerve representation | − | −180° ~ +180° | − |
| Binaural SELD | Wilkins et al. (state of the art) [64] | CRNN | Mel-spectrogram + GCC | 13 | −180° ~ +180° | − |
| | **Proposed** | CRNN | Mel-spectrogram + V-map + ITD-map + ILD-map + SC-map | 12 | −180° ~ +180° | −30° ~ +60° |

In binaural SED, recent studies have shown that replacing mono-channel input with binaural signals improves detection performance, especially in overlapping sound scenarios. Adavanne and Virtanen [51] demonstrated this using CRNN-based models trained on binaural mel-spectrograms and binaural magnitude–phase spectrograms for six sound event classes. Krause and Mesaros [52] further explored combinations of spatial features—including ILD, IPD, and GCC-PHAT—and found that incorporating binaural cues significantly enhances both SED and acoustic scene classification performance.

In binaural SSL, a variety of spatial features—such as ITD, ILD, and IPD—have been utilized as inputs to DNN, CNNs, and CRNNs. Youssef et al. [53] used ITD and ILD in a simple ANN to estimate azimuth, while Roden et al. [54] evaluated different input feature types including magnitude–phase and real–imaginary spectrograms. CNN-based approaches have shown effectiveness for both azimuth and elevation estimation using combinations of ILD, IPD, and log-magnitude inputs [55]–[57]. Zermini et al. [58] introduced a DNN with time-frequency masking, and Ma et al. [59] incorporated head movement to enhance localization in reverberant environments. Hybrid models using CNN-CRNN structures have also been proposed. Yang et al. [60] used separate branches for inter-channel magnitude and phase processing, while García-Barrios et al. [61] integrated quaternion features to compensate for head rotation. Dwivedi et al. [62] focused on elevation estimation in the median plane using spectral notches and linear prediction residuals. Notably, van der Heijden and Mehrkanoon [63] introduced neurobiologically inspired CNNs simulating human auditory nerve spatialization, achieving accurate localization in the horizontal plane.

Among deep learning-based SELD studies, the only work explicitly addressing binaural SELD is by Wilkins et al. [64]. They conducted a comparative analysis of the DCASE 2022 SELD baseline model across three input formats: FOA, binaural, and stereo. To ensure fair comparison, elevation estimation was excluded and evaluation was restricted to the horizontal plane. The results showed that although binaural and stereo inputs contain less spatial information than FOA, they still achieved reasonable localization performance for lateral sound events, supporting the viability of binaural SELD in constrained sensing scenarios.

Two-channel microphone arrays offer practical advantages for SELD systems, including low cost, reduced data complexity, easy maintenance, and compact hardware size. These properties make them well-suited for humanoid robots and telepresence applications, where system simplicity and human-like spatial hearing are essential. In particular, binaural audio replicates the way humans perceive directionality through interaural cues, enabling immersive and intuitive interpretation of sound environments. Such characteristics reinforce the viability of two-channel SELD frameworks in human-robot interaction systems.

## III. INPUT FEATURE AND DATASET FOR BISELD

### A. Binaural Sound Localization Cues of HRTFs

To design effective input features for BiSELD, we first analyze spatial cues embedded in measured head-related transfer functions (HRTFs) [65]. Psychoacoustic studies indicate that interaural time difference (ITD), interaural level difference (ILD), and spectral cues (SCs) are key to binaural sound localization [66]. In this subsection, we extract and examine these cues from an HRTF database to inform the design of binaural input representations, which are presented in the following section.

*1) Interaural Time Difference (ITD):* ITD refers to the time delay between the arrival of a sound wave at the left and right ears and is a key spatial cue for binaural sound source localization (BSSL). It is nearly zero when the sound source lies on the median plane, where the path lengths to both ears are symmetric. As the sound source deviates laterally, ITD increases with the asymmetry in propagation paths. Psychoacoustic research has shown that interaural phase delay ($ITD_P$) dominates below 1.5 kHz, while interaural envelope delay ($ITD_E$) becomes more influential above this frequency [67]. However, due to the complexity of phase unwrapping and the dependence of $ITD_E$ on signal type, direct analysis is challenging. In this study, we estimate ITD using normalized cross-correlation between left and right head-related impulse responses (HRIRs). The ITD is defined as the time delay that maximizes the similarity between the two HRIRs, computed as:

$$ITD(\theta,\phi) = \operatorname*{argmax}_{\tau} \frac{\int_{-\infty}^{+\infty} h_L(\theta,\phi,t) h_R(\theta,\phi,t-\tau)\, dt}{\sqrt{\left[\int_{-\infty}^{+\infty} h_L^2(\theta,\phi,t)\, dt\right]\left[\int_{-\infty}^{+\infty} h_R^2(\theta,\phi,t)\, dt\right]}} \quad (1)$$

with $|\tau| \leq 1000\ \mu s$,

where $\tau$ is time delay, $h_L$ and $h_R$ are the left and right HRIRs, respectively. The estimated ITD values exhibit intuitive spatial characteristics: they are approximately zero at azimuth angles of 0° and 180°, and increase as the sound source moves toward ±90°, reaching a maximum at lateral positions. Notably, near these lateral extremes, a large azimuthal shift corresponds to only a small change in ITD, indicating decreased angular resolution. In addition, ITD variations are most prominent along the horizontal plane ($\phi = 0°$) and decrease at higher or lower elevations, reflecting the geometry of human head acoustics.

*2) Interaural Level Difference (ILD):* ILD is a key localization cue, particularly effective at frequencies above 1.5 kHz. It arises from the head shadow effect, where sound reaching the ear opposite to the sound source (contralateral ear) is attenuated due to acoustic obstruction by the head. Conversely, the ear on the same side as the source (ipsilateral ear) receives a slightly amplified signal. This interaural level disparity becomes more pronounced at higher frequencies, contributing significantly to horizontal localization. In the far-field and under narrowband assumptions, ILD for a given direction is defined as:

$$ILD_{narr}(\theta,\phi,f) = 20\log_{10}\left|\frac{H_R(\theta,\phi,f)}{H_L(\theta,\phi,f)}\right|. \quad (2)$$

ILD is a multivariate function of source distance, direction, and frequency. However, under far-field conditions, ILD becomes independent of distance, depending solely on direction and frequency. In the horizontal plane, ILD is approximately zero at frontal (0°) and rear (180°) directions, and increases toward lateral (±90°) directions. At low frequencies (e.g., below 1 kHz), the head shadow effect is minimal, resulting in small and smoothly varying ILD values. In contrast, at higher frequencies (above 5 kHz), ILD magnitudes increase and exhibit more complex azimuthal variation. Notably, beyond 3.2 kHz, ILD curves become asymmetric with respect to ±90°, due to anatomical factors such as the pinna, head shape, and ear position. This front-back asymmetry in ILD serves as an additional spatial cue that can help resolve front–back confusion in binaural localization.

*3) Spectral Cue (SC):* Spectral cues (SCs), particularly those above 5 kHz, arise from the reflection and diffraction effects of the pinna, and serve as important monaural localization cues for vertical (elevation) perception [66]. Unlike ITD and ILD, SCs are encoded in the fine spectral structure of each ear's HRTF. Specifically, high-frequency notches are associated with the concha cavity [68], while spectral peaks are generated by the pinna's resonant properties [69]. Among these, the elevation-dependent frequency shift of notches is widely recognized as a key cue for elevation estimation [70]. In this study, we analyzed these spectral patterns using the measured HRTF database, focusing on the median plane. Pinna-related transfer functions (PRTFs) were extracted by applying a 2 ms Hanning window to the early part of the HRIR centered at its peak, thereby isolating the pinna-induced reflections [66]. The resulting PRTFs were Fourier-transformed, and local spectral maxima and minima were examined to track how peak and notch positions vary with elevation.

The spectral distribution of the PRTFs on the median plane reveals distinct elevation-dependent patterns. The first peak (P1), located around 4 kHz, and the second peak (P2), near 10 kHz, remain relatively stable across elevation angles. In contrast, the first notch (N1) and second notch (N2) exhibit significant frequency shifts as elevation changes. For example, N1 moves from 8 kHz to 10 kHz when the elevation varies from −40° to 90°, while N2 shows even more dramatic variation [65].

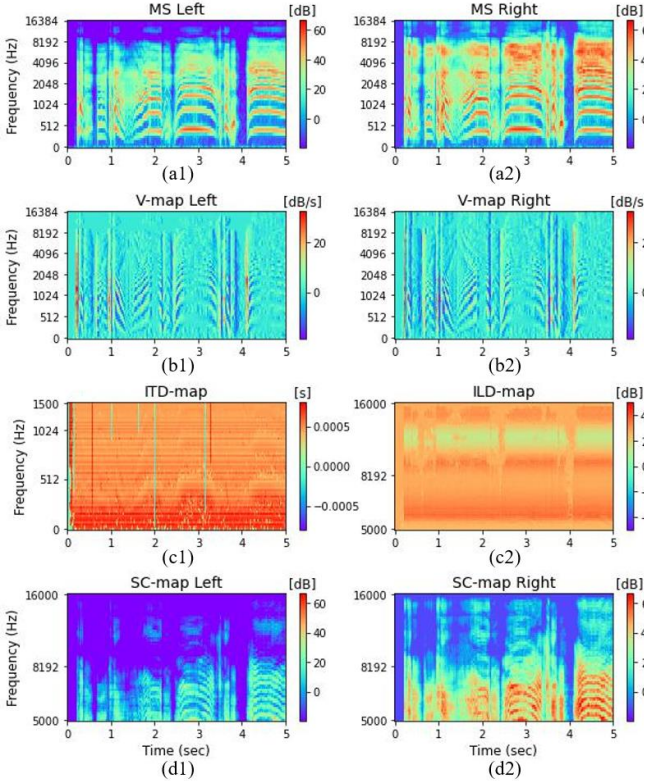

**Fig. 3.** Binaural time-frequency feature (BTFF) of a baby crying sound event from $\theta = 90°$ and $\phi = 0°$: (a1) left MS, (a2) right MS, (b1) left V-map, (b2) right V-map, (c1) ITD-map, (c2) ILD-map, (d1) left SC-map, and (d2) right SC-map.

These shifting notch frequencies are considered critical spectral cues for vertical localization. Furthermore, asymmetries in the PRTF patterns between front and back directions provide useful information to help resolve front-back confusion in binaural sound source localization. Detailed visualizations of these peak and notch patterns across elevations are available in our previous work [65].

*B. Binaural Time-Frequency Feature (BTFF)*

In deep learning, effective feature engineering aims to simplify the structure of data manifolds and enhance model robustness to input variations. For the BiSELD task, we designed an input representation that explicitly incorporates spatial cues derived from HRTFs. The proposed binaural time-frequency feature (BTFF) is an 8-channel input representation composed of domain-informed features tailored to each sub-task of BiSELD. It consists of mel-spectrogram (MS) and velocity-map (V-map) for SED; interaural-time-difference-map (ITD-map) and interaural-level-difference-map (ILD-map) for azimuth estimation; and spectral-cue-map (SC-map) for elevation estimation as shown in Fig. 3.

*1) Mel-spectrogram (MS):* The mel-spectrogram (MS) encodes the periodicity, amplitude modulation (AM), frequency modulation (FM), and onset/offset characteristics of sound events, making it well-suited for identifying their temporal boundaries. It provides a perceptually aligned time-frequency representation, where the mel scale preserves relative harmonic spacing even under pitch shifts. This property enhances the ability of convolutional layers to capture local, shift-invariant patterns, which are essential for SED. In this work, MS features are extracted separately from the left and right audio channels using short-time Fourier transform (STFT), mel-filterbank projection, and logarithmic compression.

*2) Velocity-map (V-map):* Since most sound events are transient and short-lived, it is crucial to capture the temporal dynamics of their acoustic features. The velocity-map (V-map) enhances SED by encoding the rate of change in spectrogram-based features over time [70]. Rapid temporal changes often indicate the onset or offset of sound events, allowing the model to better distinguish between acoustically similar but temporally different events. In our implementation, the V-map is computed separately for the left and right channels by applying finite differences along the time axis of the magnitude spectrogram. A combination of forward, central, and backward differences is used to account for the temporal variation at the boundaries and in the middle of the sequence.

$$V\text{-}map_{L,R}(m, b) = Mel[V_{L,R}(m, k)], \quad (3)$$

$$V_{L,R}(m, k) = \begin{cases} S_{L,R}(m+1, k) - S_{L,R}(m, k); & (m = 1) \\ \{S_{L,R}(m+1, k) - S_{L,R}(m-1, k)\}/2; & (1 < m < M) \\ S_{L,R}(m, k) - S_{L,R}(m-1, k); & (m = M) \end{cases} \quad (4)$$

where $S_{L,R}(m, k)$ and $V_{L,R}(m, k)$ are spectrogram and velocity elements at $m$ and $k$.

*3) Interaural-Time-Difference-map (ITD-map):* The ITD-map is inspired by the ITD processing mechanism of the medial superior olive (MSO) in the human brainstem, where azimuth estimation is performed by comparing neuronal firing times from both ears [71], [72]. As described in Section III-A1, ITD is a dominant localization cue below 1.5 kHz, but above this frequency, the head size becomes larger than the wavelength, causing the interaural phase difference (IPD) to exceed $2\pi$ and lose interpretability. Although IPD can theoretically be recovered through phase unwrapping, this is computationally unstable due to window truncation and spectral nulls [73], [74]. To overcome this, we propose a phase-derived ITD-map that does not require explicit phase unwrapping. The map is calculated by taking the imaginary part of the logarithmic ratio between the complex spectra of the left and right channels, which yields a phase delay that can be converted into time delay using frequency information. The final ITD-map is constructed by applying this method to frequency bins below 1.5 kHz and projecting the result onto the mel scale.

$$\Delta\tau(m, k) = \frac{1}{\omega} Im\left[ ln \frac{P_R(m, k)}{P_L(m, k)} \right] \quad (5)$$

where $P_L(m, k)$ and $P_R(m, k)$ are Fourier transform of left and right input signal respectively. Thus, the ITD-map is defined as follows:

$$ITD\text{-}map(m, b) = Mel[\Delta\tau(m, k)] \text{ for } k \leq k_{1500} \quad (6)$$

where $k_{1500}$ is the frequency index corresponding to 1.5 kHz. Therefore, the feature including ITD information can be directly extracted without phase unwrapping, using (5) and (6).

TABLE II
DATABASE COLLECTION OF SOUND EVENT AND BACKGROUND NOISE FOR BINAURAL SET CONSTRUCTION

| Database | Type | Class | Sampling Freq. (kHz) | Total (wav-files) | Length (seconds) |
|---|---|---|---|---|---|
| NIGENS [76] | Sound event | *Alarm*, *baby*, *crash*, *dog*, engine, *female scream*, *female speech*, *fire*, footsteps, general, *knock*, *male scream*, *male speech*, *phone*, piano. | 44.1 | 898 | 16,759 |
| DCASE2016-2 [77] | Sound event | Clearing throat, *cough*, door slam, drawer, keyboard, keys, *knock*, laughter, page turn, *phone*, speech. | 44.1 | 220 | 265 |

*4) Interaural-Level-Difference-map (ILD-map):* The ILD-map is motivated by the level-sensitive encoding in the lateral superior olive (LSO), where binaural intensity differences are processed through excitatory–inhibitory interactions between both ears [75]. As discussed in Section III-A2, ILD becomes a dominant localization cue above 5 kHz, where head shadow effects cause significant sound level differences between the ears. In contrast, below 1.5 kHz, sound diffraction around the head leads to minimal ILD. To capture this frequency-dependent cue, we compute the ILD-map by measuring the logarithmic power difference between the right and left channels in the frequency range above 5 kHz. The result is then projected onto the mel scale to form a time-frequency representation suitable for convolutional processing.

$$ILD\text{-}map(m,b) = Mel[\Delta S(m,k)] \text{ for } k > k_{5000} \quad (7)$$
$$\Delta S(m,k) = 10\log_{10}|P_R(m,k)/P_L(m,k)|^2, \quad (8)$$

where $\Delta S(m,k)$ is the ILD at $m$ and $k$; $k_{5000}$ is the frequency index corresponding to 5 kHz.

*5) Spectral-Cue-map (SC-map):* The SC-map is designed to capture elevation-dependent spectral notches introduced by pinna-related filtering, which play a critical role in vertical sound localization [66]. As discussed in Section III-A3, high-frequency spectral notches above 5 kHz vary systematically with the elevation of the sound source, making them a reliable monaural cue for elevation estimation. To incorporate this information, we extend the conventional mel-spectrogram by extracting and isolating the frequency bands above 5 kHz, thereby forming a high-frequency spectral map. These SC-maps are computed separately for the left and right channels and serve as dedicated inputs to capture elevation-specific spectral features. The left or right SC-map is defined as follows:

$$SC\text{-}map_{L,R}(m,b) = Mel[S_{L,R}(m,k)] \text{ for } k > k_{5000} \quad (9)$$

*C. Binaural Sound Event Dataset (Binaural Set)*

Due to the limitations in diversity, availability, and scalability of real-world sound event datasets, we constructed a synthetic dataset to train and evaluate the networks for BiSELD. Synthetic data enable precise control over event types, spatial positions, and background conditions, allowing for systematic experimentation across a wide range of acoustic scenarios. In addition to its cost-effectiveness and flexibility, data synthesis also helps avoid ethical concerns that may arise from recording in private or sensitive environments. These advantages make synthesized data particularly suitable for training SELD models with spatial awareness.

*1) Database Collection:* To construct the Binaural Set for training, validation, and testing, we synthesized each binaural sample by convolving a sound event waveform with a head-related impulse response (HRIR) corresponding to a specific direction, followed by the addition of background noise. To support this process, we collected separate databases for HRIRs, foreground sound events, and background noise, enabling flexible generation of diverse spatial audio scenes for the BiSELD task.

TABLE III
TOTAL NUMBER OF DATA AND LABEL PAIRS IN THE BINAURAL SET UNDER VARIOUS CONDITIONS

| | Train | Valid | Test | Test-H | Test-V | Total |
|---|---|---|---|---|---|---|
| Data | 672 | 144 | 144 | 36 | 12 | 1,008 |
| Length (s) | 40,320 | 8,640 | 8,640 | 2,160 | 720 | 60,480 |

First, we curated an HRTF subset from the measured KAIST HRTF database [65]. To balance directional coverage and dataset size, we selected 12 azimuth angles (−180° to +180°, at 30° intervals) and 4 elevation angles (−30° to +60°, at 30° intervals), resulting in 48 distinct spatial directions for generating binaural samples. Then, two sound event datasets were used to construct the foreground source pool, as summarized in Table II. Database 1 (NIGENS) [76] contains 714 high-quality isolated sound events across 14 strongly labeled classes with precise onset and offset annotations, making it suitable for CASA and SED tasks. Database 2 (DCASE2016 Task 2) [77] provides 30-second recordings captured in diverse real-world environments across Finland, contributing acoustic variability to the synthesized scenes.

*2) Data and Label Generation:* Each sample in the Binaural Set consists of a two-channel audio file (WAV) and its corresponding annotation file (CSV) containing sound event class, temporal boundaries, and spatial coordinates (azimuth and elevation). These annotations serve as ground truth for training and evaluation of BiSELDnet.

To preserve spectral localization cues below 16 kHz [65], all audio sources and HRIRs were resampled to 32 kHz. Each binaural sample was generated by convolving a 5-second sound event with an HRIR from one of 48 spatial directions. For each sound class, 20 samples were prepared and split into training, validation, and test sets in a 14:3:3 ratio. During synthesis, 12 binaural event samples (one from each class) were randomly selected and temporally arranged to generate a 60-second audio mixture. In the clean condition, samples were concatenated without noise addition.

As summarized in Tables III, we constructed Binaural Sets. Two additional test subsets—Test-H and Test-V—were created to assess localization performance separately in the horizontal and vertical directions.

## IV. Architecture of BiSELDnet

As discussed in Section III-B, carefully designed input features can simplify the data manifold, enhancing model robustness. According to the manifold hypothesis, high-dimensional sensory data often lies near a lower-dimensional manifold, allowing deep neural networks to learn structured mappings through geometric transformations. Leveraging this principle, we propose BiSELDnet, each tailored to effectively solve the BiSELD task using structured input features introduced in Section III.

A CRNN-based deep neural network was adopted for the BiSELD task, which demands both spectro-temporal analysis and spatial reasoning. The convolutional layers extract local time-frequency patterns from BTFF, while the recurrent layers model their temporal dynamics. This architecture is particularly suitable for variable-length sound events, as recurrent layers can process sequences of arbitrary duration—an essential requirement for real-world SELD scenarios.

The architecture of BiSELDnet, illustrated in Fig. 4(a), is a CRNN-based model designed to process the multi-channel BTFF input. The feature extractor encodes spectro-temporal and spatial cues, which are then passed through a series of convolutional modules to capture local time–frequency patterns. These modules apply convolution, normalization, activation, and max-pooling to progressively reduce the spatial resolution while preserving key features. The output of the convolutional stack is flattened into a sequence and passed to a bidirectional GRU network to model temporal context. A subsequent DNN with three fully connected layers maps each time frame to a set of 3D DOA vectors corresponding to 12 sound event classes. As shown in Fig. 4(b), each DOA vector is interpreted as a point on a unit sphere. Events are detected when the magnitude of a DOA vector exceeds 0.5v (Fig. 4c), and their spatial location is estimated from the direction of the vector (Fig. 4d).

During training, the target value for each active sound event class was set to 1, and 0 otherwise. Similarly, the corresponding 3D location vector ($x$, $y$, $z$) was used as the target for active events, while (0, 0, 0) was assigned for inactive ones [79]. BiSELDnet was trained using mean square error (MSE) loss for both detection and localization outputs. The network was trained for up to 1,000 epochs using the Adam optimizer (batch size: 128), with early stopping applied if the validation SELD error did not improve for 50 epochs.

BiSELDnet contains 763,020 parameters. All versions of BiSELDnet were implemented in Python (Keras + TensorFlow 2.5) and trained on a system with 3× NVIDIA RTX 3090 GPUs and 128 GB RAM running Ubuntu 20.04.

## V. Performance Evaluation of BTFF on BiSELDnet

### A. Evaluation Metrics

We adopt the standard evaluation metrics from the DCASE SELD task, as the BiSELD task shares the same output structure and objectives.

*1) Sound Event Detection (SED) Metrics:* We use the segment-wise F-score and error rate (ER), following the DCASE Challenge evaluation protocol [28], [77]. A sound event is considered active in a one-second segment if it appears in at least one frame within that segment. The F-score measures the harmonic mean of precision and recall, while the ER accounts for substitution, deletion, and insertion errors. An ideal SED system achieves an F-score of 1 and an ER of 0.

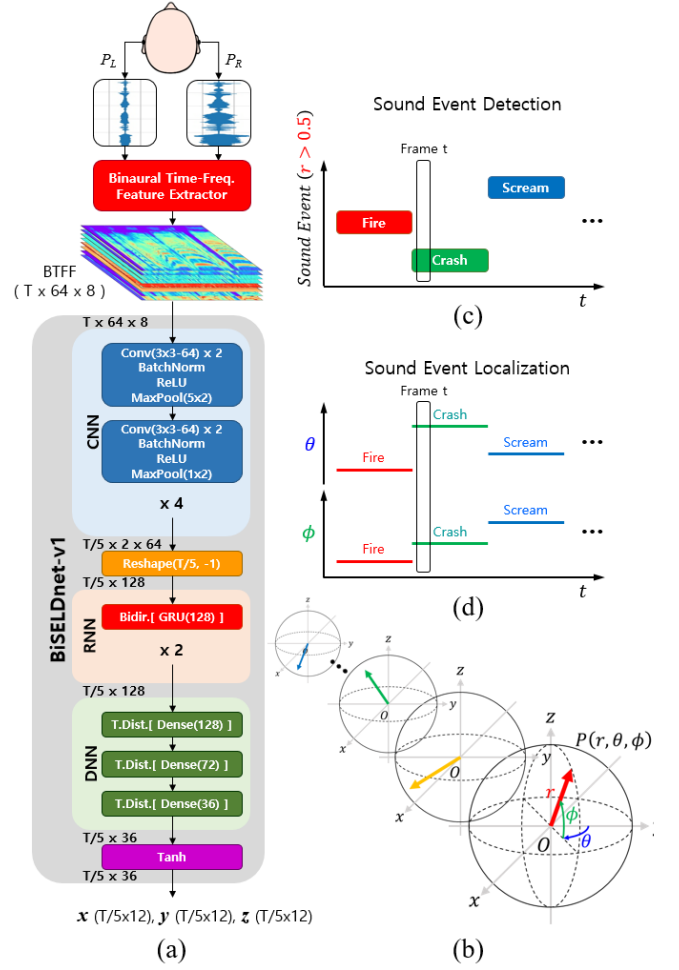

**Fig. 4.** CRNN based BiSELD model: (a) BiSELDnet architecture with BTFF, (b) coordinate system conversion of output vectors (Cartesian → spherical), (c) result of sound event detection, and (d) result of sound event localization.

*2) Direction of Arrival (DOA) Metrics:* For localization evaluation, we use localization error (LE) and localization recall (LR) as defined in the DCASE SELD framework [28]. LE measures the average angular distance between the reference and predicted DOA vectors, while LR indicates the proportion of correctly localized sound events. LE ranges from 0° (perfect localization) to 180°, and LR from 0 (no matches) to 1 (perfect recall).

*3) Sound Event Localization and Detection (SELD) Metrics:* For a comprehensive evaluation of the BiSELD task, we adopt four joint SELD metrics following [80]: location-aware detection metrics ($ER_{20°}$, $F_{20°}$) and class-aware localization metrics ($LE_{CD}$, $LR_{CD}$). $ER_{20°}$ and $F_{20°}$ consider a detection correct only if the predicted class matches the reference and the angular error is less than 20°. $LE_{CD}$ and $LR_{CD}$ assess localization accuracy for correctly predicted classes, regardless of angular threshold. To provide a single scalar performance index, the SELD error is computed as the average of SED and DOA errors, each normalized between 0 and 1:

$$SELD\ error = \frac{SED\ error + DOA\ error}{2} \quad (10)$$

where the SED error is defined as follows:

$$SED\ error = \frac{ER_{20°} + (1 - F_{20°})}{2} \quad (11)$$

and the DOA error is defined as follows:

$$DOA\ error = \frac{LE_{CD}/180 + (1 - LR_{CD})}{2} \quad (12)$$

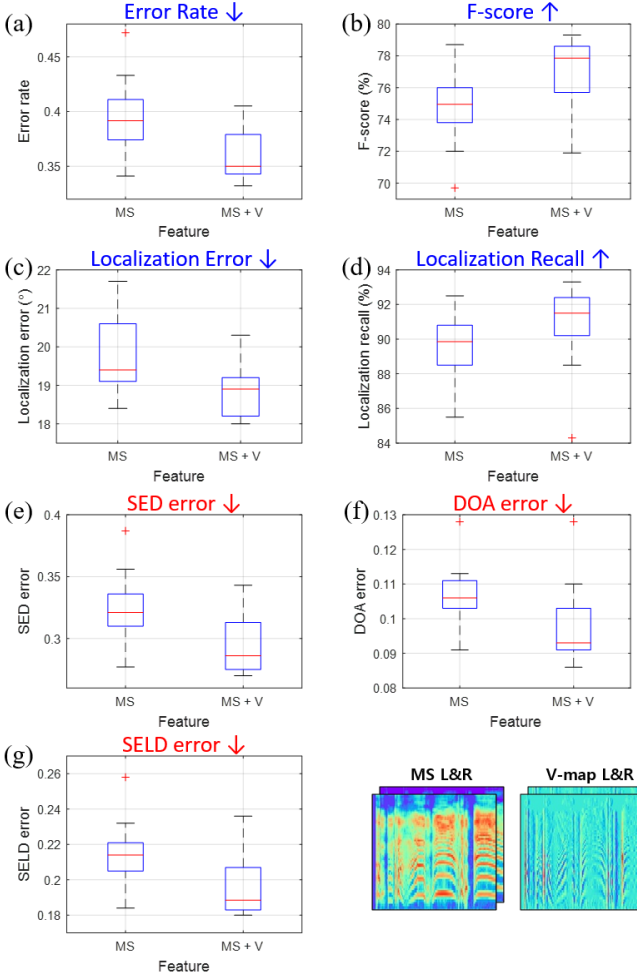

**Fig. 5.** Evaluation results of BiSELDnet on test set with MS and MS + V-map (MS + V) as input features.

An ideal model will have a SELD error, SED error and DOA error of zero.

### B. Performance Evaluation of BTFF

To assess the contribution of each BTFF sub-feature, ablation experiments were conducted using the test sets (Table III) and the BiSELDnet described in Section VI. MS (mel-spectrogram) served as the baseline feature. To evaluate detection performance, V-map was concatenated to MS. For horizontal localization, ITD-map and ILD-map were added to MS. For vertical localization, SC-map was combined with MS. Finally, to validate the effectiveness of the full BTFF representation, all five sub-features—MS, V-map, ITD-map, ILD-map, and SC-map—were concatenated and evaluated using the same setting. The evaluation confirmed the performance gains from jointly leveraging spectro-temporal patterns and HRTF-based spatial cues. Each configuration was trained and evaluated ten times to ensure statistical reliability.

TABLE IV
MEDIAN VALUES FOR BiSELDNET PERFORMANCE WITH MS
AND MS + V-MAP AS INPUT FEATURES ON TEST SET.

|  | Detection | | | Localization | | | Total |
|---|---|---|---|---|---|---|---|
|  | $ER_{20°}\downarrow$ | $F_{20°}\uparrow$ (%) | SED Error↓ | $LE_{CD}\downarrow$ (°) | $LR_{CD}\uparrow$ (%) | DOA Error↓ | SELD Error↓ |
| MS | 0.392 | 75.0 | 0.321 | 19.4 | 89.9 | 0.106 | 0.214 |
| MS +V | **0.350** | **77.9** | **0.286** | **18.9** | **91.5** | **0.093** | **0.189** |

TABLE V
MEDIAN VALUES FOR BiSELDNET PERFORMANCE WITH
MS, MS + ITD-MAP, AND MS + ITD-MAP + ILD-MAP AS
INPUT FEATURES ON TEST SET H

|  | Detection | | | Localization | | | Total |
|---|---|---|---|---|---|---|---|
|  | $ER_{20°}\downarrow$ | $F_{20°}\uparrow$ (%) | SED Error↓ | $LE_{CD}\downarrow$ (°) | $LR_{CD}\uparrow$ (%) | DOA Error↓ | SELD Error↓ |
| MS | 0.387 | 75.0 | 0.319 | 17.3 | 87.4 | 0.112 | 0.217 |
| MS +ITD | 0.278 | 81.6 | 0.236 | 11.9 | 90.1 | 0.081 | 0.162 |
| MS +ITD +ILD | **0.235** | **85.8** | **0.187** | **4.2** | **90.8** | **0.060** | **0.124** |

*1) Combination of MS and V-map:* To investigate the impact of V-map on binaural sound event detection, we compared the performance of BiSELDnet trained with MS alone versus MS concatenated with V-map (denoted as MS + V). As shown in Fig. 5 and summarized in Table IV, MS + V achieved a lower error rate (ER) of 0.350 compared to 0.392, and a higher F-score of 77.9% compared to 75.0%, reducing the SED error from 0.321 to 0.286. These results demonstrate that incorporating V-map effectively enhances detection performance. Additionally, due to the interdependence between detection and localization metrics in the SELD task, MS + V also showed a slight improvement in localization, reducing LE from 19.4° to 18.9° and increasing LR from 89.9% to 91.5%, which led to a decrease in DOA error from 0.106 to 0.093.

Incorporating V-map as an additional input feature allows models to capture temporal dynamics and transitions in sound events more effectively [70]. By representing the rate of change in spectro-temporal features, V-map helps distinguish variations in pitch, timbre, loudness, and tempo—crucial for accurate sound event detection. Furthermore, by emphasizing frame-wise differences, V-map improves robustness to noise and environmental interference, making the model more resilient under real-world acoustic conditions. This enhancement contributes to the improved detection performance observed in the BiSELD task.

*2) Combination of MS, ITD-map, and ILD-map on the Horizontal Plane:* To evaluate the effect of ITD-map and ILD-map on horizontal localization, BiSELDnet was trained with different input combinations using the Test-H dataset from

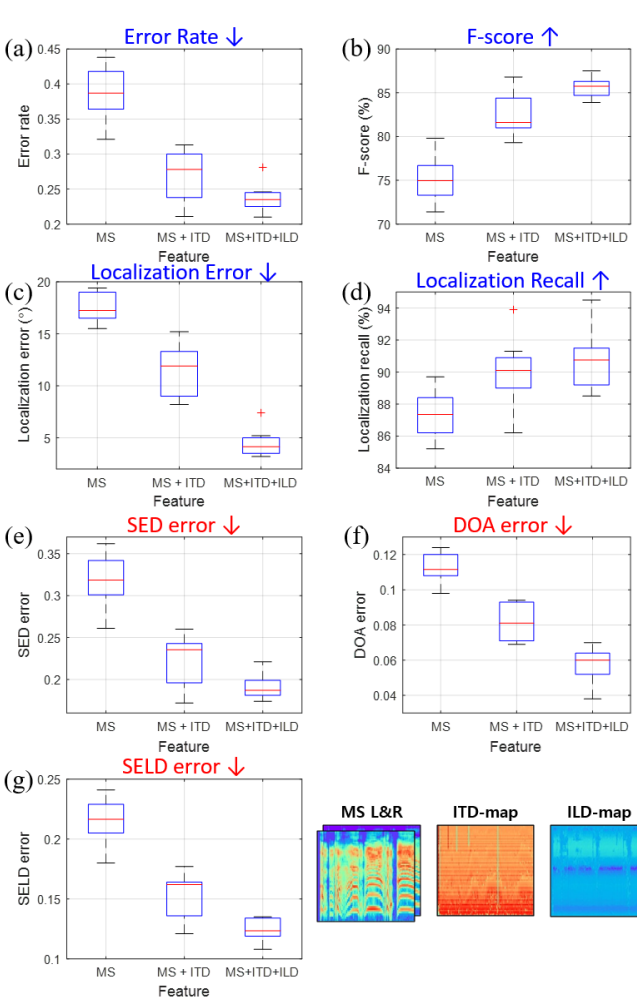

**Fig. 6.** Evaluation results of BiSELDnet on test set H with MS, MS + ITD-map (MS + ITD), and MS + ITD-map + ILD-map (MS + ITD + ILD) as input features.

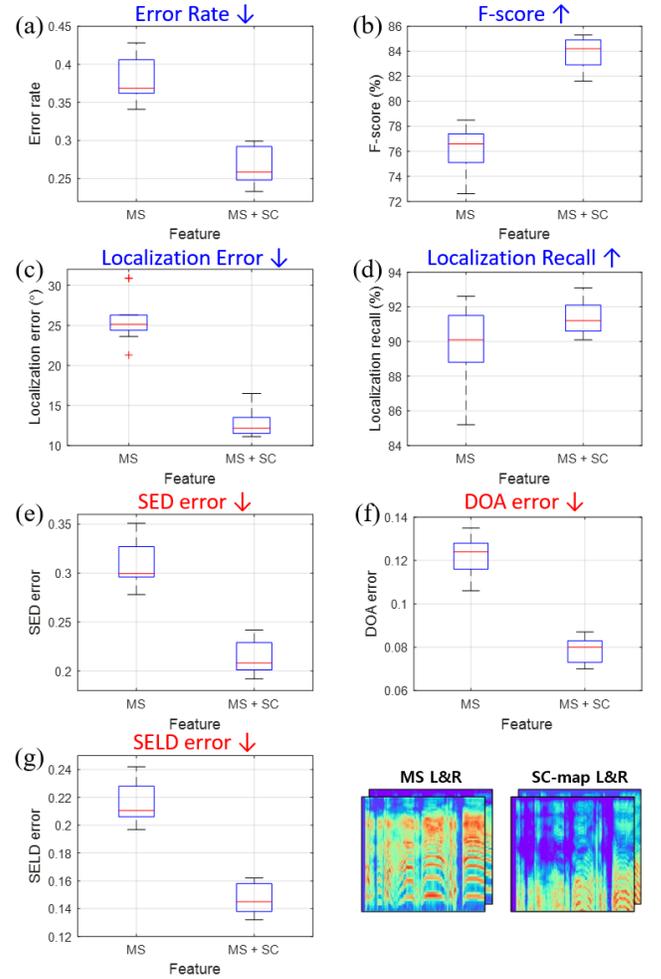

**Fig. 7.** Evaluation results of BiSELDnet on test set V with MS and MS + SC-map (MS + SC) as input features.

Table III. The results are shown in Fig. 6 and summarized in Table V. Compared to using only MS, the addition of ITD-map (MS + ITD) significantly reduced the localization error (LE) from 17.3° to 11.9°, and increased localization recall (LR) from 87.4% to 90.1%, resulting in a decrease in DOA error from 0.112 to 0.081. Further inclusion of ILD-map (MS + ITD + ILD) reduced LE to 4.2°, improved LR to 90.8%, and lowered DOA error to 0.060. These results confirm the effectiveness of combining ITD and ILD features for accurate azimuth estimation.

This also translated to better detection performance. Compared to the baseline, SED error decreased from 0.319 to 0.236 with MS + ITD, and to 0.187 with MS + ITD + ILD, accompanied by corresponding improvements in ER (from 0.387 to 0.235) and F-score (from 75.0% to 85.8%). These improvements highlight the interdependency between detection and localization in BiSELD, and validate the complementary role of ILD-map in refining azimuth estimation, as qualitatively discussed in Section V-B.

When localizing sound events on the horizontal plane, BiSELDnet leverages ITD-map and ILD-map as complementary sub-features. The ITD-map captures interaural time differences (ITDs) that arise due to the sound reaching each ear at slightly different times, particularly effective below 1.5 kHz where the wavelength exceeds head width and phase ambiguity is minimal. In contrast, ILD-map encodes interaural level differences (ILDs), which become prominent above 5 kHz due to the head shadow effect. While ITD is dominant at low frequencies, ILD provides reliable spatial cues at high frequencies. By jointly learning from these frequency-specific localization cues, BiSELDnet achieves robust azimuth estimation across the audible spectrum.

ITD-map and ILD-map offer complementary localization cues across different frequency ranges—ITD being effective at low frequencies and ILD at high frequencies. Notably, ILD-map also mitigates front-back confusion inherent in ITD-based azimuth estimation. While ITD patterns exhibit symmetry between front and back, ILD patterns vary asymmetrically at high frequencies due to head and pinna shape. This front-back asymmetry in ILD-map provides additional cues for disambiguating azimuth, enhancing localization accuracy on the horizontal plane.

*3) Combination of MS and SC-map on the Median Plane:* To assess the effect of SC-map on vertical localization, the performance of BiSELDnet trained with MS was compared to that with MS + SC using the Test-V dataset. The results are shown in Fig. 7 and summarized in Table VI. When SC-map was added, the localization error (LE) decreased from 25.2° to 12.2°, and localization recall (LR) improved from 90.1% to 91.2%, reducing DOA error from 0.124 to 0.080. These results confirm that SC-map effectively captures vertical localization cues on the median plane, consistent with prior analyses of high-frequency spectral notches caused by the pinna. Furthermore, the sound event detection performance also improved: error rate (ER) dropped from 0.369 to 0.259, and F-score increased from 76.6% to 84.2%, resulting in a lower SED error from 0.300 to 0.208. This suggests that enhanced vertical localization also contributes to better detection performance in the BiSELD task.

TABLE VI
MEDIAN VALUES FOR BISELDNET PERFORMANCE WITH MS AND MS + SC-MAP AS INPUT FEATURES ON TEST SET V

|  | Detection | | | Localization | | | Total |
|---|---|---|---|---|---|---|---|
|  | $ER_{20°}\downarrow$ | $F_{20°}\uparrow$ (%) | SED Error$\downarrow$ | $LE_{CD}\downarrow$ (°) | $LR_{CD}\uparrow$ (%) | DOA Error$\downarrow$ | SELD Error$\downarrow$ |
| MS | 0.369 | 76.6 | 0.300 | 25.2 | 90.1 | 0.124 | 0.211 |
| MS+SC | **0.259** | **84.2** | **0.208** | **12.2** | **91.2** | **0.080** | **0.145** |

TABLE VII
BEST PERFORMANCES OF BISELDNET ON TEST SET.

|  | Detection | | | Localization | | | Total |
|---|---|---|---|---|---|---|---|
|  | $ER_{20°}\downarrow$ | $F_{20°}\uparrow$ (%) | SED Error$\downarrow$ | $LE_{CD}\downarrow$ (°) | $LR_{CD}\uparrow$ (%) | DOA Error$\downarrow$ | SELD Error$\downarrow$ |
| BTFF | 0.210 | 87.1 | 0.169 | 4.4 | 92.1 | 0.052 | 0.110 |

Spectral cues above 5 kHz play a critical role in vertical localization by encoding elevation-specific information. These high-frequency cues result from the filtering effects of the pinna and head, producing characteristic notch patterns that vary with elevation. As analyzed in the HRTF-based SC study, the first notch (N1) appears around 8 kHz and systematically shifts to higher frequencies as the elevation increases. It is therefore inferred that BiSELDnet learns such frequency-elevation dependencies from the SC-map, particularly the elevation-dependent movement of N1.

*4) Performance of final BTFF on BiSELDnet:*

To evaluate the full performance of the proposed BTFF, BiSELDnet was trained and tested using the complete BTFF feature set. Table VII summarizes the best performance of BiSELDnet on the test set. The model achieved an error rate $ER_{20°}$ of 0.210 and an F-score $F_{20°}$ of 87.1%, resulting in a low SED error of 0.169. For localization, the model achieved a class-dependent localization error $LE_{CD}$ of 4.4°, with a localization recall $LR_{CD}$ of 92.1%, yielding a DOA error of 0.052. Consequently, the final SELD error was 0.110. These results confirm that the proposed BTFF enables BiSELDnet to effectively perform joint detection and localization of binaural sound events with high accuracy across both spatial and temporal domains.

## VI. CONCLUSION

To emulate human auditory perception in intelligent machines, this study proposed Binaural Sound Event Localization and Detection (BiSELD) as a new task and introduced a structured framework to address it. A synthetic dataset, the Binaural Set, was constructed using measured HRTFs and diverse acoustic conditions to enable controlled training and evaluation. To effectively learn spatial hearing cues, a novel input representation, the Binaural Time-Frequency Feature (BTFF), was designed based on psychoacoustic insights. BTFF encodes key localization cues—interaural time difference (ITD), interaural level difference (ILD), and spectral cues (SC)—into an eight-channel time-frequency structure derived from binaural signals. The proposed BiSELDnet, a compact CRNN-based model with depthwise separable convolutions, successfully learns both spectro-temporal and spatial representations from BTFF. Evaluation results confirmed that each sub-feature contributes to performance in its respective domain: V-map enhances detection via temporal dynamics, ITD-/ILD-maps improve horizontal localization through complementary frequency-domain cues, and SC-map enables accurate vertical localization by capturing pinna-related spectral patterns. Collectively, these components allow BiSELDnet to achieve accurate joint detection and localization, demonstrating that integrating HRTF-based spatial cues into a deep learning framework is a promising approach to replicating human-like auditory perception in humanoid systems.

REFERENCES

[1] S. Saeedvand, M. Jafari, H. S. Aghdasi, and J. Baltes, "A comprehensive survey on humanoid robot development," *Knowl. Eng. Rev.*, vol. 34, pp. 1–18, Dec. 2019, doi: 10.1017/S02698889190001582.

[2] J. Englsberger et al., "Overview of the torque-controlled humanoid robot TORO," in *Proc. 14th IEEE-RAS Int. Conf. Humanoid Robots*, Madrid, Spain, Nov. 2014, pp. 916–923.

[3] J.-I. Yamaguchi, A. Takanishi, and I. Kato, "Development of a biped walking robot compensating for three-axis moment by trunk motion," in *Proc. IEEE/RSJ Int. Conf. Intelligent Robots and Systems (IROS)*, Yokohama, Japan, Jul. 1993, pp. 561–566.

[4] K. Hirai, M. Hirose, Y. Haikawa, and T. Takenaka, "The development of Honda humanoid robot," in *Proc. IEEE Int. Conf. Robotics and Automation (ICRA)*, Leuven, Belgium, May 1998, pp. 1321–1326.

[5] M. Gienger, K. Löffler, and F. Pfeiffer, "Towards the design of a biped jogging robot," in *Proc. IEEE Int. Conf. Robotics and Automation (ICRA)*, Seoul, Korea, May 2001, pp. 4140–4145.

[6] K. Kaneko et al., "Design of prototype humanoid robotics platform for HRP," in *Proc. IEEE/RSJ Int. Conf. Intelligent Robots and Systems (IROS)*, Lausanne, Switzerland, Dec. 2002, pp. 2431–2436.

[7] Y. Sakagami, R. Watanabe, C. Aoyama, S. Matsunaga, N. Higaki, and K. Fujimura, "The intelligent ASIMO: System overview and integration," in *Proc. IEEE/RSJ Int. Conf. Intelligent Robots and Systems (IROS)*, Lausanne, Switzerland, Dec. 2002, pp. 2478–2483.

[8] I.-W. Park, J.-Y. Kim, J. Lee, and J.-H. Oh, "Mechanical design of humanoid robot platform KHR-3 (KAIST humanoid robot - 3: HUBO)," in *Proc. 5th IEEE-RAS Int. Conf. Humanoid Robots (ICHR)*, Tsukuba, Japan, Dec. 2005, pp. 321–326.

[9] J.-H. Oh, D. Hanson, W.-S. Kim, I.-Y. Han, J.-Y. Kim, and I.-W. Park, "Design of android type humanoid robot Albert HUBO," in *Proc. IEEE/RSJ Int. Conf. Intelligent Robots and Systems (IROS)*, Beijing, China, Oct. 2006, pp. 1428–1433.

[10] J. A. Rojas-Quintero and M. C. Rodríguez-Liñán, "A literature review of sensor heads for humanoid robots," *Robot. Auton. Syst.*, vol. 143, pp. 1–21, Sep. 2021, doi: 10.1016/j.robot.2021.103834.

[11] N. J. Ferrier, "The Harvard binocular head," in *Proc. SPIE 1708, Applications of Artificial Intelligence X: Machine Vision and Robotics*, Orlando, FL, USA, Mar. 1992, pp. 1–13.

[12] I. Kato *et al.*, "The robot musician 'wabot-2' (waseda robot-2)," *Robotics*, vol. 3, no. 2, pp. 143–155, Jun. 1987, doi: 10.1016/0167-8493(87)90002-7.

[13] R. A. Russell and A. H. Purnamadjaja, "Odor and airflow: Complementary senses for a humanoid robot," in *Proc. IEEE Int. Conf. Robotics and Automation (ICRA)*, Washington, DC, USA, May 2002, pp. 1842–1847.

[14] M. Bonnefille, "Electronic nose technology and applications," in *Practical Analysis of Flavor and Fragrance Materials*, K. Goodner and R. Rouseff, Eds. Chichester, UK: Wiley, 2011, pp. 111–154.

[15] L. Natale, G. Metta, and G. Sandini, "Development of auditory-evoked reflexes: Visuo-acoustic cues integration in a binocular head," *Robot. Auton. Syst.*, vol. 39, no. 2, pp. 87–106, May 2002, doi: 10.1016/S0921-8890(02)00174-4.

[16] G.-T. Lee, "Binaural sound event localization and detection neural network based on HRTF localization cues for humanoid robots," Ph.D. Dissertation, KAIST, 2024

[17] A. S. Bregman, *Auditory Scene Analysis. The Perceptual Organization of Sound.* Cambridge, MA, USA: MIT Press, 1990, pp. 3–9.

[18] D. Wang and G. J. Brown, "Fundamentals of computational auditory scene analysis," in *Computational Auditory Scene Analysis: Principles, Algorithms, and Applications*, D. Wang and G. J. Brown, Eds. Piscataway, NJ, USA: Wiley-IEEE Press, 2006, pp. 11–14.

[19] B. T. Szabó, S. L. Denham, and I. Winkler, "Computational models of auditory scene analysis: A review," *Front. Neurosci.*, vol. 10, pp. 1–16, Nov. 2016, doi: 10.3389/fnins.2016.00524.

[20] D. Barniv and I. Nelken, "Auditory streaming as an online classification process with evidence accumulation," *PLoS One*, vol. 10, no. 12, pp. 1–20, Dec. 2015, doi: 10.1371/journal.pone.0144788.

[21] J. Nix and V. Hohmann, "Combined estimation of spectral envelopes and sound source direction of concurrent voices by multidimensional statistical filtering," *IEEE Trans. Audio Speech Lang. Process.*, vol. 15, no. 3, pp. 995–1008, Mar. 2007, doi: 10.1109/TASL.2006.889788.

[22] D. Wang and P. Chang, "An oscillatory correlation model of auditory streaming," *Cogn. Neurodynamics*, vol. 2, no. 1, pp. 7–19, Mar. 2008, doi: 10.1007/s11571-007-9035-8.

[23] R. Pichevar and J. Rouat, "Monophonic sound source separation with an unsupervised network of spiking neurons," *Neurocomputing*, vol. 71, no. 1-3, pp. 109–120, Dec. 2007, doi: 10.1016/j.neucom.2007.08.001.

[24] R. W. Mill, T. M. Böhm, A. Bendixen, I. Winkler, and S. L. Denham, "Modelling the emergence and dynamics of perceptual organisation in auditory streaming," *PLoS Comput. Biol.*, vol. 9, no. 3, pp. 1–21, Mar. 2013, doi: 10.1371/journal.pcbi.1002925.

[25] J. Rankin, E. Sussman, and J. Rinzel, "Neuromechanistic model of auditory bistability," *PLoS Comput. Biol.*, vol. 11, no. 11, pp. 1–34, Nov. 2015, doi: 10.1371/journal.pcbi.1004555.

[26] L. Krishnan, M. Elhilali, and S. Shamma, "Segregating complex sound sources through temporal coherence," *PLoS Comput. Biol.*, vol. 10, no. 12, pp. 1–10, Dec. 2014, doi: 10.1371/journal.pcbi.1003985.

[27] M. Elhilali and S. A. Shamma, "A cocktail party with a cortical twist: How cortical mechanisms contribute to sound segregation," *J. Acoust. Soc. Am.*, vol. 124, no. 6, pp. 3751–3771, Dec. 2008, doi: 10.1121/1.3001672.

[28] S. Adavanne, A. Politis, J. Nikunen, and T. Virtanen, "Sound event localization and detection of overlapping sources using convolutional recurrent neural networks," *IEEE J. Sel. Top. Signal Process.*, vol. 13, no. 1, pp. 34–48, Mar. 2019, doi: 10.1109/JSTSP.2018.2885636.

[29] R. Takeda and K. Komatani, "Sound source localization based on deep neural networks with directional activate function exploiting phase information," in *Proc. IEEE Int. Conf. Acoust., Speech, Signal Process. (ICASSP)*, Shanghai, China, Mar. 2016, pp. 405–409.

[30] N. Yalta, K. Nakadai, and T. Ogata, "Sound source localization using deep learning models," *J. Robot. Mechatron.*, vol. 29, no. 1, pp. 37–48, Feb. 2017, doi: 10.20965/jrm.2017.p0037.

[31] W. He, P. Motlicek, and J.-M. Odobez, "Deep neural networks for multiple speaker detection and localization," in *Proc. IEEE Int. Conf. Robotics and Automation (ICRA)*, Brisbane, Australia, May 2018, pp. 74–79.

[32] M. Crocco, M. Cristani, A. Trucco, and V. Murino, "Audio surveillance: A systematic review," *ACM Comput. Surv.*, vol. 48, no. 4, pp. 1–46, Feb. 2016, doi: 10.1145/2871183.

[33] G. Valenzise, L. Gerosa, M. Tagliasacchi, F. Antonacci, and A. Sarti, "Scream and gunshot detection and localization for audio-surveillance systems," in *Proc. IEEE Conf. Advanced Video and Signal Based Surveillance (AVSS)*, London, UK, Sep. 2007, pp. 21–26.

[34] S. Chu, S. Narayanan, and C.-C. J. Kuo, "Environmental sound recognition with time–frequency audio features," *IEEE Trans. Audio Speech Lang. Process.*, vol. 17, no. 6, pp. 1142–1158, Aug. 2009, doi: 10.1109/TASL.2009.2017438.

[35] H. Sun, X. Liu, K. Xu, J. Miao, and Q. Luo, "Emergency vehicles audio detection and localization in autonomous driving," 2021, *arXiv:2109.14797*.

[36] A. Politis, A. Mesaros, S. Adavanne, T. Heittola, and T. Virtanen, "Overview and evaluation of sound event localization and detection in DCASE 2019," *IEEE-ACM Trans. Audio Speech Lang.*, vol. 29, pp. 684–698, Dec. 2020, doi: 10.1109/TASLP.2020.3047233.

[37] Z. Lu, "Sound event detection and localization based on CNN and LSTM," Detection Classification Acoust. Scenes Events (DCASE) Challenge, Tech. Rep., pp. 1–3, 2019.

[38] Q. Li, X. Zhang, and H. Li, "Online direction of arrival estimation based on deep learning," in *Proc. IEEE Int. Conf. Acoust., Speech, Signal Process. (ICASSP)*, Calgary, AB, Canada, Apr. 2018, pp. 2616–2620.

[39] A. Bohlender, A. Spriet, W. Tirry, and N. Madhu, "Exploiting temporal context in CNN based multisource DOA estimation," *IEEE/ACM Trans. Audio Speech Lang. Process.*, vol. 29, no. 1, pp. 1594–1608, Mar. 2021, doi: 10.1109/TASLP.2021.3067113.

[40] F. Ronchini, D. Arteaga, and A. Pérez-López, "Sound event localization and detection based on CRNN using rectangular filters and channel rotation data augmentation," in *Proc. Detection Classification Acoust. Scenes Events Workshop (DCASE Workshop)*, Tokyo, Japan, Nov. 2020, pp. 180–184.

[41] H. A. C. Maruri, P. L. Meyer, J. Huang, JAdH. Ontiveros, and H. Lu, "GCC-PHAT cross-correlation audio features for simultaneous sound event localization and detection (SELD) in multiple rooms," Detection Classification Acoust. Scenes Events (DCASE) Challenge, Tech. Rep., pp. 1–4, 2019.

[42] Y. Cao *et al.*, "Polyphonic sound event detection and localization using a two-stage strategy," in *Proc. Detection Classification Acoust. Scenes Events Workshop (DCASE Workshop)*, New York, NY, USA, Oct. 2019, pp. 30–34.

[43] Y. Cao *et al.*, "Two-stage sound event localization and detection using intensity vector and generalized cross-correlation," Detection Classification Acoust. Scenes Events (DCASE) Challenge, Tech. Rep., pp. 1–4, 2019.

[44] A. Sampathkumar and D. Kowerko, "Sound event detection and localization using CRNN models," Detection Classification Acoust. Scenes Events (DCASE) Challenge, Tech. Rep., pp. 1–3, 2020.

[45] Y. Cao, T. Iqbal, Q. Kong, F. An, W. Wang, and M. D. Plumbley, "An improved event-independent network for polyphonic sound event localization and detection," in *Proc. IEEE Int. Conf. Acoust., Speech, Signal Process. (ICASSP)*, Toronto, ON, Canada, Jun. 2021, pp. 885–889.

[46] T. N. T. Nguyen *et al.*, "A general network architecture for sound event localization and detection using transfer learning and recurrent neural network," in *Proc. IEEE Int. Conf. Acoust., Speech, Signal Process. (ICASSP)*, Toronto, ON, Canada, Jun. 2021, pp. 935–939.

[47] K. Shimada, Y. Koyama, N. Takahashi, S. Takahashi, and Y. Mitsufuji, "ACCDOA: Activity-coupled Cartesian direction of arrival representation for sound event localization and detection," in *Proc. IEEE Int. Conf. Acoust., Speech, Signal Process. (ICASSP)*, Toronto, ON, Canada, Jun. 2021, pp. 915–919.

[48] K. Shimada, Y. Koyama, S. Takahashi, N. Takahashi, E. Tsunoo, and Y. Mitsufuji, "Multi-ACCDOA: Localizing and detecting overlapping sounds from the same class with auxiliary duplicating permutation invariant training," in *Proc. IEEE Int. Conf. Acoust., Speech, Signal Process. (ICASSP)*, Singapore, Singapore, May 2022, pp. 316–320.

[49] T. N. T. Nguyen, K. N. Watcharasupat, N. K. Nguyen, D. L. Jones, and W.-S. Gan, "SALSA: Spatial cue-augmented log-spectrogram features for polyphonic sound event localization and detection," *IEEE-ACM Trans. Audio Speech Lang.*, vol. 30, pp. 1749–1762, May 2022, doi: 10.1109/TASLP.2022.3173054.

[50] T. N. T. Nguyen, D. L. Jones, K. N. Watcharasupat, H. Phan, and W.-S. Gan, "SALSA-Lite: A fast and effective feature for polyphonic sound event localization and detection with microphone arrays," in *Proc. IEEE Int. Conf. Acoust., Speech, Signal Process. (ICASSP)*, Singapore, Singapore, May 2022, pp. 716–720.

[51] S. Adavanne and T. Virtanen, "A report on sound event detection with different binaural features," in *Proc. Detection Classification Acoust. Scenes Events Workshop (DCASE Workshop)*, Munich, Germany, Nov. 2017, pp. 1–5.

[52] D. A. Krause and A. Mesaros, "Binaural signal representations for joint sound event detection and acoustic scene classification," in *Proc. 30th Euro. Signal Process. Conf. (EUSIPCO)*, Belgrade, Serbia, Aug. 2022, pp. 399–403.
[53] K. Youssef, S. Argentieri, and J.-L. Zarader, "A learning-based approach to robust binaural sound localization," in *Proc. IEEE/RSJ Int. Conf. Intelligent Robots and Systems (IROS)*, Tokyo, Japan, Nov. 2013, pp. 2927–2932.
[54] R. Roden, N. Moritz, S. Gerlach, S. Weinzierl, and S. Goetze, "On sound source localization of speech signals using deep neural networks," in *Proc. Deutsche Jahrestagung Akustik (DAGA)*, Nuremberg, Germany, Mar. 2015, pp. 1510–1513.
[55] Q. V. Nguyen, L. Girin, G. Bailly, F. Elisei, and D. C. Nguyen, "Autonomous sensorimotor learning for sound source localization by a humanoid robot," in *Proc. IEEE/RSJ Int. Conf. Intelligent Robots and Systems (IROS)*, Madrid, Spain, Oct. 2018, pp. 1–4.
[56] C. Pang, H. Liu, and X. Li, "Multitask learning of time-frequency CNN for sound source localization," *IEEE Access*, vol. 7, pp. 40725–40737, Mar. 2019, doi: 10.1109/ACCESS.2019.2905617.
[57] Y. Yang, J. Xi, W. Zhang, and L. Zhang, "Full-sphere binaural sound source localization using multi-task neural network," in *Proc. Asia-Pacific Signal and Information Processing Association Annual Summit and Conference (APSIPA ASC)*, Auckland, New Zealand, Dec. 2020, pp. 432–436.
[58] A. Zermini, Y. Yu, Y. Xu, M. D. Plumbley, and W. Wang, "Deep neural network based audio source separation," in *Proc. Int. Conf. Math. Signal Proc. (IMA)*, Birmingham, UK, May 2016, pp. 1–4.
[59] N. Ma, T. May, and G. J. Brown, "Exploiting deep neural networks and head movements for robust binaural localization of multiple sources in reverberant environments," *IEEE/ACM Trans. Audio Speech Lang. Process.*, vol. 25, no. 12, pp. 2444–2453, Dec. 2017, doi: 10.1109/TASLP.2017.2750760.
[60] B. Yang, H. Liu, and X. Li, "Learning deep direct-path relative transfer function for binaural sound source localization," *IEEE/ACM Trans. Audio Speech Lang. Process.*, vol. 29, pp. 3491–3503, Oct. 2021, doi: 10.1109/TASLP.2021.3120641.
[61] G. García-Barrios, D. A. Krause, A. Politis, A. Mesaros, J. M. Gutiérrez-Arriola, and R. Fraile, "Binaural source localization using deep learning and head rotation information," in *Proc. 30th Euro. Signal Process. Conf. (EUSIPCO)*, Belgrade, Serbia, Aug. 2022, pp. 36–40.
[62] P. Dwivedi, G. Routray, and R. M. Hegde, "Binaural source localization in median plane using learning based method for robot audition," in *Proc. 24th Int. Congr. Acoust. (ICA)*, Gyeongju, Korea, Oct. 2022, pp. 1–8.
[63] K. van der Heijden and S. Mehrkanoon, "Goal-driven, neurobiological-inspired convolutional neural network models of human spatial hearing," *Neurocomputing*, vol. 470, pp. 432–442, Jan. 2022, doi: 10.1016/j.neucom.2021.05.104.
[64] J. Wilkins, M. Fuentes, L. Bondi, S. Ghaffarzadegan, A. Abavisani, and J. P. Bello, "Two vs. four-channel sound event localization and detection," in *Proc. Detection Classification Acoust. Scenes Events Workshop (DCASE Workshop)*, Tampere, Finland, Sep. 2023, pp. 1–5.
[65] G.-T. Lee, S.-M. Choi, B.-Y. Ko, and Y.-H. Park, "HRTF measurement for accurate sound localization cues," 2022, *arXiv:2203.03166v2*.
[66] K. Iida, *Head-Related Transfer Function and Acoustic Virtual Reality*. Narashino, Japan: Springer, 2019, pp. 15–55.
[67] B. Xie, *Head-Related Transfer Function and Virtual Auditory Display*, 2nd ed., Plantation, FL, USA: J. Ross Publishing, 2013, pp. 81–85.
[68] E. A. G. Shaw and R. Teranishi, "Sound pressure generated in an external-ear replica and real human ears by a nearby point source," *J. Acoust. Soc. Am.*, vol. 44, no. 1, pp. 240–249, Jul. 1968, doi: 10.1121/1.1911059.
[69] H. Takemoto, P. Mokhtari, H. Kato, R. Nishimura, and K. Iida, "Mechanism for generating peaks and notches of head-related transfer functions in the median plane," *J. Acoust. Soc. Am.*, vol. 132, no. 6, pp. 3832–3841, Dec. 2012, doi: 10.1121/1.4765083.
[70] G.-T. Lee, H. Nam, S.-H. Kim, S.-M. Choi, Y. Kim, and Y.-H. Park, "Deep learning based cough detection camera using enhanced features," *Expert Syst. Appl.*, vol. 206, pp. 1–20, Nov. 2022, doi: 10.1016/j.eswa.2022.117811.
[71] G.-T. Lee and Y.-H. Park, "Estimation of interaural time difference based on cochlear filter bank and ZCPA auditory model," *Trans. Korean Soc. Noise Vib. Eng.*, vol. 29, no. 6, pp. 722–734, Dec. 2019, doi: 10.5050/KSNVE.2019.29.6.722.
[72] G.-T. Lee and Y.-H. Park, "Method for estimating interaural time differences based on cochlea filter bank and EUZ auditory model in noisy environment," in *Proc. Int. Conf. Noise Control Eng. (Inter-noise)*, Seoul, Korea, Aug. 2020, pp. 1–12.
[73] R. M. Hegde, H. A. Murthy, and V. R. R. Gadde, "Significance of the modified group delay feature in speech recognition," *IEEE Trans. Audio Speech Lang. Process.*, vol. 15, no. 1, pp. 190–202, Jan. 2007, doi: 10.1109/TASL.2006.876858.
[74] B. Yegnanarayana, "Group delay spectrogram of speech signals without phase wrapping," *J. Acoust. Soc. Am.*, vol. 151, no. 3, pp. 2181–2191, Mar. 2022, doi: 10.1121/10.0009922.
[75] R. F. Lyon, *Human and Machine Hearing: Extracting Meaning from Sound*. Cambridge, UK: Cambridge Univ. Press, 2017, pp. 66–67.
[76] I. Trowitzsch, J. Taghia, Y. Kashef, and K. Obermayer, "The NIGENS general sound events database," 2020, *arXiv:1902.08314*.
[77] A. Mesaros *et al.*, "Detection and classification of acoustic scenes and events: Outcome of the DCASE 2016 challenge," *IEEE-ACM Trans. Audio Speech Lang.*, vol. 26, no. 2, pp. 379–393, Feb. 2018, doi: 10.1109/TASLP.2017.2778423.
[78] A. Mesaros, T. Heittola, and T. Virtanen, "A multi-device dataset for urban acoustic scene classification," in *Proc. Detection Classification Acoust. Scenes Events Workshop (DCASE Workshop)*, Surrey, UK, Nov. 2018, pp. 1–5.
[79] G.-T. Lee and Y.-H. Park, "Binaural sound event localization and detection for humanoid robot," in *Proc. 24th Int. Congr. Acoust. (ICA)*, Gyeongju, Korea, Oct. 2022, pp. 1–12.
[80] A. Mesaros, S. Adavanne, A. Politis, T. Heittola, and T. Virtanen, "Joint measurement of localization and detection of sound events," in *Proc. IEEE Workshop Appl. Signal Process. Audio Acoust. (WASPAA)*, New Paltz, NY, USA, Oct. 2019, pp. 333–337.